\documentclass[aps,twocolumn,showpacs,preprintnumbers,amsmath,amssymb,pra]{revtex4}
\usepackage{graphicx} %,amssymb,amstext,amsmath,amsthm,
\usepackage{pstricks}
\usepackage{pstricks-add}
\usepackage{color}
\begin{document}

% Use the \preprint command to place your local institutional report
% number in the upper righthand corner of the title page in preprint mode.
% Multiple \preprint commands are allowed.
% Use the 'preprintnumbers' class option to override journal defaults
% to display numbers if necessary
%\preprint{}

%Title of paper
\title{Reduced fidelity in topological quantum phase transitions}

% repeat the \author .. \affiliation  etc. as needed
% \email, \thanks, \homepage, \altaffiliation all apply to the current
% author. Explanatory text should go in the []'s, actual e-mail
% address or url should go in the {}'s for \email and \homepage.
% Please use the appropriate macro foreach each type of information

% \affiliation command applies to all authors since the last
% \affiliation command. The \affiliation command should follow the
% other information
% \affiliation can be followed by \email, \homepage, \thanks as well.
\author{Erik Eriksson}
%\thanks{}
%\email[]{erik.eriksson@physics.gu.se}
\author{Henrik Johannesson}
%\email[]{henrik.johannesson@physics.gu.se}
%\homepage[]{Your web page}
%\thanks{}
%\altaffiliation{}
\affiliation{Department of Physics, University of Gothenburg, SE 412 96 Gothenburg,
Sweden}

%Collaboration name if desired (requires use of superscriptaddress
%option in \documentclass). \noaffiliation is required (may also be
%used with the \author command).
%\collaboration can be followed by \email, \homepage, \thanks as well.
%\collaboration{}
%\noaffiliation

%\date{\today}

\begin{abstract}
We study the reduced fidelity between local states of lattice systems exhibiting topological order. By exploiting mappings to spin models with classical order, we are able to analytically extract the scaling behavior of the reduced fidelity at the corresponding quantum phase transitions out of the topologically ordered phases. Our results suggest that the reduced fidelity, albeit being a local measure, generically serves as an accurate marker of a topological quantum phase transition.
\end{abstract}
% insert suggested PACS numbers in braces on next line
\pacs{03.67.-a, 64.70.Tg, 03.65.Vf}
% insert suggested keywords - APS authors don't need to do this
%\keywords{}

%\maketitle must follow title, authors, abstract, \pacs, and \keywords
\maketitle

% body of paper here - Use proper section commands
% References should be done using the \cite, \ref, and \label commands
%\section{}
% Put \label in argument of \section for cross-referencing
%\section{\label{}}
%\subsection{}
%\subsubsection{}

%\section{Introduction}
{\em Introduction $-$} Electron correlations in condensed matter systems sometimes produce 
topologically ordered phases where effects from local perturbations are exponentially suppressed~\cite{WenReview}. The most prominent examples are the fifty or so observed fractional quantum Hall phases, with the topological order manifested in gapless edge states and excitations with fractional statistics.
The fact that the ground state degeneracy in phases with non-Abelian
statistics cannot be lifted by local perturbations lies at the heart of current proposals for topological quantum computation \cite{Nayak}. 

\indent The insensitivity to local perturbations invalidates the use of a local order parameter to identify a quantum phase transition out of a topologically ordered phase. Attempts to build a theory of topological quantum phase transitions (TQPTs) $-$ replacing the Ginzburg-Landau symmetry-breaking paradigm $-$ have instead borrowed concepts from quantum information theory, in particular those of {\em entanglement entropy} \cite{PreskillWen} and {\em fidelity} \cite{hamma}, none of which require the construction of an order parameter. 

\indent Fidelity measures the similarity between two quantum states, and, for pure states, is defined as the modulus of their overlap. Since the ground state changes rapidly at a quantum phase transition, one expects that the fidelity between two ground states that differ by a small change in the driving parameter should exhibit a sharp drop. This expectation has been confirmed in a number of case studies \cite{Gu}, including several TQPTs \cite{hamma,abasto,fidelTQPT}.

\indent Suppose that one replaces the two ground states in a fidelity analysis by two states that also differ slightly in the driving parameter, but which describe only a local region of the system of interest.  The proper concept that encodes the similarity between such mixed states is that of the {\em reduced fidelity}, which is the maximum pure state overlap between purifications of the mixed states~\cite{uhlmann}. It has proven useful in the analysis of a number of ordinary symmetry-breaking quantum phase transitions~\cite{fidelQPT}. But since the reduced fidelity is a local property of the system, similarly to that of a local order parameter, one may think that it would be less sensitive  to a TQPT, which involves a global rearrangement of nonlocal quantum correlations \cite{WenReview}. However, this intuition turns out to be wrong. As we show in this paper, several TQPTs are accurately signaled by a singularity in the second-order derivative of the reduced fidelity. Moreover, the singularity can be even stronger than for the (pure state) global fidelity. The fact that a TQPT gets imprinted in a local quantity may at first seem surprising, but, as we shall see, parallels and extends results from earlier studies \cite{castelnovochamon,trebst}. 

%\section{Fidelity and fidelity susceptibility}
{\em Fidelity and fidelity susceptibility $-$}
The fidelity $F(\beta, \beta')$ between two states described by the density matrices
$\hat{\rho}(\beta)$ and $\hat{\rho}(\beta')$ is defined as~\cite{uhlmann}
\begin{equation} \label{rf}
 F(\beta,\beta') = \textrm{Tr}
\sqrt{\sqrt{\hat{\rho}(\beta)}\hat{\rho}(\beta') \sqrt{\hat{\rho}(\beta)}}.
\end{equation}
When a system is in a pure state, $\hat{\rho}(\beta) =
|\Psi(\beta)\rangle \langle\Psi(\beta)|$, $F(\beta, \beta')$ becomes just the
state overlap $|\langle \Psi(\beta')|\Psi(\beta)\rangle|$. When the states under consideration describe a subsystem, they will generally be mixed
states, and we call the fidelity between such states \textit{reduced
fidelity}.  In the limit where $\beta$ and $\beta'=\beta + \delta\beta$ are very close, it is useful to
define the {\em fidelity susceptibility}~\cite{you}
\begin{equation} \label{rfs}
\chi_F = \displaystyle \lim_{\delta\beta \to 0} \frac{-2 \ln F}{\delta\beta^2},
\end{equation}
consistent with the pure state expansion $F \!\approx \!1 - \chi_F \delta\beta^2 / 2$.  

%\section{The Castelnovo-Chamon Model}
{\em The Castelnovo-Chamon Model $-$}
The first model we consider was introduced by Castelnovo and
Chamon~\cite{castelnovochamon}, and is a deformation of the Kitaev toric code
model~\cite{kitaevtoric}. The Hamiltonian for $N$ spin-1/2 particles on the bonds of a
square lattice with periodic boundary conditions is
\begin{equation} \label{cchamiltonian}
H=-\lambda_0 \displaystyle \sum_p B_p -\lambda_1 \sum_s A_s + \lambda_1 \sum_s
e^{-\beta\sum_{i\in s}\hat{\sigma}^{z}_i},
\end{equation}
where $A_s=\prod_{i\in s} \hat{\sigma}^{x}_i$ and $B_p=\prod_{i\in p}
\hat{\sigma}^{z}_i$ are the star and plaquette operators of the original Kitaev
toric code model. The star operator $A_s$ acts on the spins around the vertex $s$,
and the plaquette operator $B_p$ acts on the spins on the boundary of the plaquette
$p$. For $\lambda_{0,1}>0$ the ground state in the topological sector containing the
fully magnetized state $|0\rangle$ is given by~\cite{castelnovochamon}
\begin{equation} \label{ccgs}
|GS(\beta)\rangle = \displaystyle \sum_{g \in G} \frac{e^{\beta\sum_i
\sigma^z_i(g)/2}}{\sqrt{Z(\beta)}}g|0\rangle ,
\end{equation}
with
$Z(\beta) =  \sum_{g \in G} e^{\beta\sum_i \sigma^z_i(g)}$,
where $G$ is the Abelian group generated by the star operators $A_s$,
and $\sigma^z_i(g)$ is the $z$ component of the spin at site $i$ in the state
$g|0\rangle$. 
When $\beta=0$ the state in (\ref{ccgs}) reduces to the topologically ordered ground state of the toric
code model~\cite{kitaevtoric}. When $\beta \to \infty$ the
ground state (\ref{ccgs}) becomes the magnetically ordered state $|0\rangle$. 
At $\beta_c= (1/2)\ln(\sqrt{2}+1)$ there is a second-order TQPT
where the topological entanglement entropy
$S_{topo}$ goes from $S_{topo}=1$ for $\beta < \beta_c$ to $S_{topo}=0$ for $\beta >
\beta_c$~\cite{castelnovochamon}. The global fidelity susceptibility $\chi_F$ close to 
$\beta_c$ was obtained
in Ref. \onlinecite{abasto}, and found to diverge as
\begin{equation} \label{globalF}
\chi_F \sim  \ln|\beta_c/\beta - 1|.
\end{equation}

\begin{figure}
\centering
	 \includegraphics[width=0.35\textwidth]{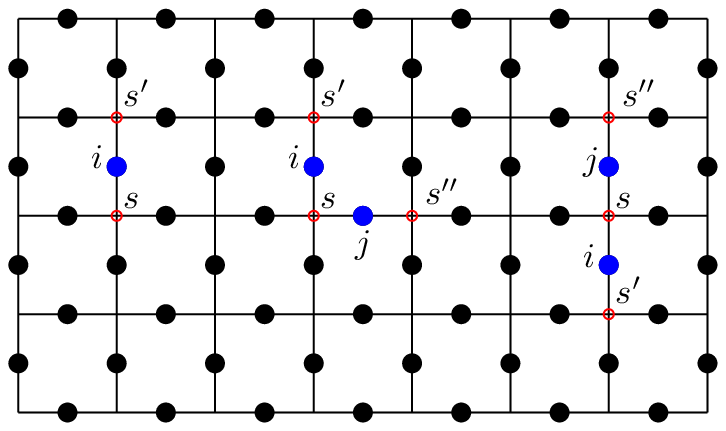}
\caption{(Color online.) Mapping between the Castelnovo-Chamon
model and the 2D Ising model. The spins of the former reside on the lattice bonds
(filled black circles), and the spins of the latter on the vertices.
Left: $\sigma_i^z=\theta_s \theta_{s'}$, where $i$ is the bond between the
neighboring vertices $\langle s,s' \rangle$. Middle and right: For $i$ and $j$
nearest (next-nearest) neighbors, the mapping gives $\langle \hat{\sigma}_i^{z}
\hat{\sigma}_j^{z} \rangle = \langle \theta_s \theta_{s'} \theta_{s''}
\theta_{s}\rangle = \langle  \theta_{s'} \theta_{s''}\rangle$, where $\langle
s',s'' \rangle$ are next-nearest (third-nearest) neighbors.}
\label{fig:cclattice}
\end{figure}

\begin{figure}[t!]
        \centering
                \includegraphics[width=0.40\textwidth]{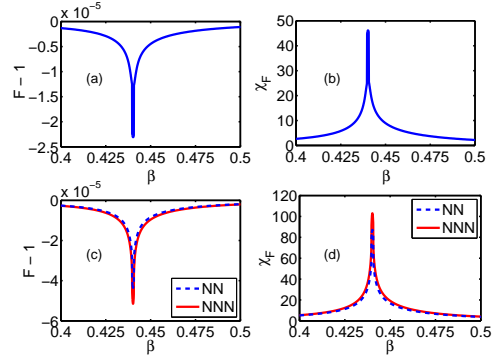}
        \caption{(Color online.) Single-site fidelity (a), single-site fidelity
susceptibility (b), two-site fidelity (c) and two-site fidelity susceptibility (d)
of the Castelnovo-Chamon model calculated with a parameter difference $\delta\beta
= 0.001$ and with $N \to \infty$. The reduced fidelity
susceptibilities will diverge according to Eq. (\ref{logdivbeta}) when $\delta\beta
\to 0$. In (c) and (d) we plot for both nearest (NN) and next-nearest (NNN) neighbors.} 
        \label{fig:ccfig}
\end{figure}

\indent We here calculate the single-site reduced fidelity between the ground states
of a single spin at two different parameter values $\beta$ and $\beta'$. To
construct the density matrix $\hat{\rho}_i$ for the spin at site $i$ we use
the expansion
\begin{equation} \label{rdm}
\hat{\rho}_i = \frac{1}{2} \displaystyle \sum_{\mu=0}^3 \langle \hat{\sigma}_i^{\mu}
\rangle \hat{\sigma}_i^{\mu},
\end{equation}
with $\hat{\sigma}_i^{0} \equiv \openone_i$, and with the expectation values taken with respect to the ground state in (\ref{ccgs}). There is a one-to-two
mapping between the configurations $\{g\}=G$ and the configurations
$\{\theta\}  \equiv \Theta$ of the classical 2D Ising model $H =
-J\sum_{<s,s'>}\theta_s\theta_{s'}$ with $\theta_s = -1 \ (+1)$ when the
corresponding star operator $A_s$ is (is not) acting on the site
$s$~\cite{castelnovochamon}. Thus $\sigma_i^z=\theta_s \theta_{s'}$, where $i$ is
the bond between the neighboring vertices $\langle s,s' \rangle$, see
Fig.~\ref{fig:cclattice}. This gives $\langle GS(\beta)| \hat{\sigma}_i^z
|GS(\beta)\rangle = (1/Z(\beta))\sum_{\theta \in \Theta} \theta_s \theta_{s'}
e^{\beta\sum_{\langle s'', s''' \rangle} \theta_{s''}\theta_{s'''}} =
 E(\beta)/N$, where $\beta$ is identified as the reduced nearest-neighbor
coupling $J/T\!=\!\beta$ of the Ising model with energy $E(\beta)$. The
two expectation values $\langle GS(\beta)| \hat{\sigma}_i^x|GS(\beta) \rangle$ and
$\langle GS(\beta)| \hat{\sigma}_i^y|GS(\beta) \rangle$ are both zero, since
$\langle 0 | g \hat{\sigma}_i^x g' |0\rangle = 0$, $\forall g,g' \in G$, and
similarly for $\hat{\sigma}_i^y$. 
It follows that $\hat{\rho}_i = (1/2)\mbox{diag}\left(1+E(\beta)/N, 1-E(\beta)/N\right)$ in the $\hat{\sigma}_i^z$ eigenbasis.
Since the density matrices at different parameter values $\beta$ and
$\beta'$ commute, the reduced fidelity (\ref{rf}) is
\begin{eqnarray} \label{fideig}
F(\beta,\beta') =
  \textrm{Tr} \sqrt{\hat{\rho}_i(\beta) \hat{\rho}_i(\beta')} =
\sum_i \sqrt{\lambda_i \lambda'_i},
\end{eqnarray}
where $\{\lambda_i\}$ ($\{\lambda'_i\}$) are the eigenvalues of
$\hat{\rho}_i(\beta)$ ($\hat{\rho}_i(\beta')$).
The energy $E(\beta)$ of the 2D Ising model in the thermodynamic limit $N\to \infty$
is given by $E(\beta)/N=-\coth(2\beta)\left[1+(2/\pi)(2\tanh^2(2\beta)-1)K(\kappa)\right]/2$, where $K(\kappa) = \int_0^{\pi/2} d\theta (1-\kappa^2 \sin^2\theta)^{-1/2}$ and
$\kappa = 2\sinh (2\beta)\, / \cosh^2 (2\beta)$~\cite{onsager}.
This gives us the plot of the single-site fidelity shown in Fig.~\ref{fig:ccfig},
where we see that the TQPT at
$\beta_c=(1/2)\ln(\sqrt{2}+1) \approx 0.44$ is marked by a sudden drop in the
fidelity. \\
\indent The single-site fidelity susceptibility $\chi_F$ is
\begin{equation} \label{rfsder}
\chi_F =  \sum_i \frac{(\partial_{\beta} \lambda_i)^2}{4\lambda_i},
\end{equation}
for commuting density matrices~\cite{xiong}. Here $\partial_{\beta} \lambda_{1,2}\!=\!\pm
(2N)^{-1}\partial_{\beta}E(\beta)\!=\!\pm (2N\beta^2)^{-1} C(\beta)$, with $C(\beta)$ the specific heat of the 2D Ising model. Thus $\chi_F$ diverges as
\begin{equation}  \label{logdivbeta}
 \chi_F \sim \ln^2 |\beta_c / \beta -1|,
\end{equation}
at $\beta_c$, faster than the global fidelity susceptibility in (\ref{globalF}). In Fig. \ref{fig:ccfig} we plot the single-site fidelity
susceptibility using Eq. (\ref{rfs}), but with finite $\delta\beta
= 0.001$.

\indent The two-site fidelity can be obtained in a similar way. We expand the reduced density
matrix $\hat{\rho}_{ij}$ as
\begin{equation} \label{rdm2}
\hat{\rho}_{ij} = \frac{1}{4} \displaystyle \sum_{\mu,\nu=0}^3 \langle
\hat{\sigma}_i^{\mu} \hat{\sigma}_j^{\nu}\rangle \hat{\sigma}_i^{\mu}
\hat{\sigma}_j^{\nu}.
\end{equation}
The only non-zero expectation values in (\ref{rdm2}) are $\langle
\hat{\sigma}_i^{0}\hat{\sigma}_j^{0}\rangle=1$, $\langle
\hat{\sigma}_i^{z}\hat{\sigma}_j^{0}\rangle= \langle \hat{\sigma}_i^{z}\rangle$,
$\langle \hat{\sigma}_i^{0}\hat{\sigma}_j^{z}\rangle= \langle
\hat{\sigma}_j^{z}\rangle$ and $\langle
\hat{\sigma}_i^{z}\hat{\sigma}_j^{z}\rangle$.  Translational invariance 
implies that $\langle \hat{\sigma}_j^{z}\rangle = \langle
\hat{\sigma}_i^{z}\rangle$, so that
\begin{equation} \label{rdm2cc}
\hat{\rho}_{ij} = \frac{1}{4} ( 1 +  \langle \hat{\sigma}_i^{z}\rangle (
\hat{\sigma}_i^{z} +  \hat{\sigma}_j^{z})  
+ \langle \hat{\sigma}_i^{z}\hat{\sigma}_j^{z}\rangle \hat{\sigma}_i^{z}
\hat{\sigma}_j^{z} ).
\end{equation}
The
eigenvalues are seen to be $\lambda_{1,2} = (1/4)(1\pm 2\langle \hat{\sigma}_i^{z}\rangle + \langle
\hat{\sigma}_i^{z} \hat{\sigma}_j^{z} \rangle)$ and $\lambda_{3,4} = (1/4)(1- \langle \hat{\sigma}_i^{z} \hat{\sigma}_j^{z} \rangle)$. Now the fidelity can be calculated using Eq. (\ref{fideig}). Here we focus on the cases
where $i$ and $j$ are nearest and next-nearest neighbors. Then the mapping to the
2D Ising model gives that $\langle \hat{\sigma}_i^{z}\rangle = \langle
\theta_s \theta_{s'}\rangle$ and $\langle \hat{\sigma}_j^{z}\rangle = \langle
\theta_{s''} \theta_{s'''}\rangle$, where $i$ ($j$) is the bond between the
neighboring vertices $\langle s,s' \rangle$ ($\langle s'',s''' \rangle$). When $i$
and $j$ are nearest (next-nearest) neighbors, we get $\langle \hat{\sigma}_i^{z}
\hat{\sigma}_j^{z} \rangle = \langle \theta_s \theta_{s'} \theta_{s''}
\theta_{s}\rangle = \langle  \theta_{s'} \theta_{s''}\rangle$, where $\langle s',s''
\rangle$ are next-nearest (third-nearest) neighbors on the square lattice (cf.
Fig.~\ref{fig:cclattice}). As before, $\langle \hat{\sigma}_i^{z}\rangle =
E(\beta)/N$. We obtain $\langle  \theta_{s'}
\theta_{s''}\rangle$ from the equivalence between the 2D
Ising model and the quantum 1D XY model
\begin{equation}
H_{XY} =  - \displaystyle \sum_n (
\alpha_{+}\hat{\sigma}_n^{x}\hat{\sigma}_{n+1}^{x} +
\alpha_{-}\hat{\sigma}_n^{y}\hat{\sigma}_{n+1}^{y} +
h\hat{\sigma}_{n}^{z}),
\end{equation}
where $\alpha_{\pm} = (1\pm\gamma)/2$.
This has been shown to give
\begin{equation}
\langle \theta_{0,0}\theta_{n,n} \rangle = \langle
\hat{\sigma}_0^{x}\hat{\sigma}_n^{x} \rangle_{XY} |_{\gamma=1,h=(\sinh
2\beta)^{-2}}
\end{equation}
for Ising spins on the same diagonal, and
\begin{multline}
\langle \theta_{n,m}\theta_{n,m'} \rangle = \cosh^2 (\beta^*)\langle
\hat{\sigma}_m^{x}\hat{\sigma}_{m'}^{x} \rangle_{XY}|_{\gamma =
\gamma_{\beta},h=h_{\beta}}  \\
- \sinh^2 (\beta^*)\langle \hat{\sigma}_m^{y}\hat{\sigma}_{m'}^{y}
\rangle_{XY}|_{\gamma = \gamma_{\beta},h=h_{\beta}} 
\end{multline}
for Ising spins on the same row (or, by symmetry, column), where $\tanh \beta^* =
e^{-2\beta}$, $\gamma_{\beta} = (\cosh 2\beta^*)^{-1}$ and $h_{\beta}=
(1-\gamma^2)^{1/2} / \tanh 2\beta$~\cite{suzuki}. Known results for the 1D XY model give~\cite{barouchmccoy}
\begin{eqnarray} \label{corrxy}
\langle \hat{\sigma}_m^{x}\hat{\sigma}_{m+r}^{x} \rangle_{XY} &=& \left|
\begin{array}{cccc}
G_{-1} & G_{-2} & \ldots & G_{-r}  \\
G_0 & G_{-1} & \ldots & G_{-r+1}\\
\vdots & \vdots & \ddots & \vdots \\
G_{r-2} & G_{r-3} & \ldots & G_{-1}
\end{array} \right|, \\
\langle \hat{\sigma}_m^{y}\hat{\sigma}_{m+r}^{y} \rangle_{XY} &=& \left|
\begin{array}{cccc}
G_{1} & G_{0} & \ldots & G_{-r+2}  \\
G_2 & G_{1} & \ldots & G_{-r+3}\\
\vdots & \vdots & \ddots & \vdots \\
G_{r} & G_{r-1} & \ldots & G_{1}
\end{array} \right|,
\end{eqnarray}
where
$ G_{r'} =(1/ \pi ) \int_0^{\pi} d\phi \, (h-\cos \phi) \cos
(\phi r') / \Lambda_{\phi}(h)
+ (\gamma / \pi) \int_0^{\pi} d\phi \, \sin \phi \sin (\phi
r') / \Lambda_{\phi}(h)$
and $\Lambda_{\phi}(h) = ((\gamma \sin \phi)^2 + (h-\cos \phi)^2)^{1/2}$. These
relations allow us to plot the two-site fidelity, and also the two-site fidelity
susceptibility using Eq. (\ref{rfs}), see Fig.~\ref{fig:ccfig}. Note that the two-site functions are only
slightly different depending 
on whether the two sites are nearest neighbors or next-nearest neighbors. It follows from Eq.~(\ref{rfsder}) that also the two-site $\chi_F$ has a stronger divergence at criticality than the global fidelity susceptibility.\\
\indent It is interesting to note the slight asymmetry of the reduced fidelities around the critical point, seen in Fig.~\ref{fig:ccfig}, indicating a somewhat smaller response to changes in the driving parameter in the topological phase.

%\section{The transverse Wen-plaquette model}
{\em The transverse Wen-plaquette model $-$}
We now turn to the transverse Wen-plaquette model, obtained from the ordinary Wen-plaquette
model~\cite{wenplaquette} for spin-1/2 particles on the vertices of a square lattice
by adding a magnetic field $h$~\cite{transversewen},
\begin{equation} \label{twp}
H= g\sum_i \hat{F}_i + h \sum_i \hat{\sigma}_i^x ,
\end{equation}
where $\hat{F}_i = \hat{\sigma}_i^x  \hat{\sigma}_{i+\hat{x}}^y
\hat{\sigma}_{i+\hat{x}+\hat{y}}^x \hat{\sigma}_{i+\hat{y}}^y $ and $g<0$. The
boundary conditions are periodic. At $h=0$ the ground state is the topologically
ordered ground state of the Wen-plaquette model~\cite{wenplaquette} and in the limit
$h \to \infty$ the ground state is magnetically ordered. Since $\hat{F}_i$,
$\hat{\sigma}_j^x$ have the same commutation relations as
$\hat{\tau}_{i+\hat{x}/2+\hat{y}/2}^z$, $\hat{\tau}_{j-\hat{x}/2+\hat{y}/2}^x
\hat{\tau}_{j+\hat{x}/2-\hat{y}/2}^x$ (where the Pauli matrices $\hat{\tau}$ act on
spin-1/2 particles at the centers of the plaquettes), the Hamiltonian (\ref{twp})
can be mapped onto independent quantum Ising chains,
\begin{equation} \label{twpi}
H= -h \sum_a \sum_i \left( g_I  \hat{\tau}_{a,i+\frac{1}{2}}^z +
\hat{\tau}_{a,i-\frac{1}{2}}^x \hat{\tau}_{a,i+\frac{1}{2}}^x \right),
\end{equation}
with $g_I = g/h$, and where $\hat{\tau}_{i+\frac{1}{2}}^z$ and $ \hat{\tau}_{i-\frac{1}{2}}^x \hat{\tau}_{i+\frac{1}{2}}^x$ are the images of $\hat{\sigma}_i^x  \hat{\sigma}_{i+\hat{x}}^y \hat{\sigma}_{i+\hat{x}+\hat{y}}^x
\hat{\sigma}_{i+\hat{y}}^y$ and $\hat{\sigma}_i^x$ respectively \cite{transversewen}. The index $a$ denotes the diagonal chains over the plaquette-centered sites,
and $i$ is the site index on each diagonal chain.
Known results for criticality in the quantum Ising chain imply that the transverse Wen-plaquette model has a
TQPT at $g/h=1$ \cite{transversewen}.

\indent We now calculate the reduced fidelity. The mapping onto the quantum Ising chains immediately gives that $\langle
\hat{\sigma}_{i}^x \rangle = \langle \hat{\tau}_{i-\frac{1}{2}}^x
\hat{\tau}_{i+\frac{1}{2}}^x \rangle$. In the $\hat{\sigma}_{i}^x$ basis, the
Hamiltonian (\ref{twp}) only flips spins in pairs, therefore we get $\langle
\hat{\sigma}_{i}^y \rangle = 0$ and $\langle \hat{\sigma}_{i}^z \rangle = 0$. The
single-site reduced density matrix (\ref{rdm}) is therefore given by $\hat{\rho}_i =
(1/2)( 1 +  \langle \hat{\tau}_{i-\frac{1}{2}}^x \hat{\tau}_{i+\frac{1}{2}}^x
\rangle \hat{\sigma}_{i}^x )$, which is diagonal in the $\hat{\sigma}_{i}^x$ basis,
with eigenvalues $\lambda_{1,2} = (1/2)(1 \pm \langle \hat{\tau}_{i-\frac{1}{2}}^x
\hat{\tau}_{i+\frac{1}{2}}^x \rangle )$. The single-site fidelity is thus given by
Eq. (\ref{fideig}), and $\langle \hat{\tau}_{i-\frac{1}{2}}^x
\hat{\tau}_{i+\frac{1}{2}}^x \rangle$ is calculated using Eq. (\ref{corrxy}) with
$\gamma=1$ and $h = g_I$. The result reveals that the TQPT is accompanied 
by a sudden drop in
the single-site fidelity. Now, $\partial_{g_I} \lambda_{1,2} = \pm
\frac{1}{2} \partial_{g_I} \langle \hat{\tau}_{i-\frac{1}{2}}^x
\hat{\tau}_{i+\frac{1}{2}}^x \rangle$, which diverges logarithmically at the
critical point $g_I = 1$. Therefore Eq. (\ref{rfsder}) implies that at $h/g=1$,
$\chi_F$ diverges as
\begin{equation} \label{logdiv}
\chi_F \sim \ln^2 |g/h -1|,
\end{equation}
as in Eq. (\ref{logdivbeta}) for the Castelnovo-Chamon model.

\indent We can also calculate the two-site fidelity for two nearest neighbor spins at
sites $i,j$. All non-trivial expectation values in the
expansion (\ref{rdm2}) of the reduced density matrix, except $\langle
\hat{\sigma}_i^{x}\rangle$, $ \langle \hat{\sigma}_j^{x}\rangle$ and $\langle
\hat{\sigma}_i^{x}\hat{\sigma}_j^{x}\rangle$, will be zero, since only these
operators can be constructed from those in the Hamiltonian (\ref{twp}). The mapping onto the quantum Ising chains gives $\langle \hat{\sigma}_{i}^x \hat{\sigma}_{j}^x \rangle \!=\!
\langle \hat{\tau}_{i-\frac{1}{2}}^x \hat{\tau}_{i+\frac{1}{2}}^x
\hat{\tau}_{j-\frac{1}{2}}^x \hat{\tau}_{j+\frac{1}{2}}^x\rangle \!=\! (\langle
\hat{\tau}_{i-\frac{1}{2}}^x \hat{\tau}_{i+\frac{1}{2}}^x \rangle )^2$. Thus the two-site density matrix is given by $\hat{\rho}_{ij} = (1/4) ( 1 +  \langle
\hat{\tau}_{i-\frac{1}{2}}^x \hat{\tau}_{i+\frac{1}{2}}^x \rangle
(\hat{\sigma}_{i}^x + \hat{\sigma}_{j}^x ) + (\langle \hat{\tau}_{i-\frac{1}{2}}^x
\hat{\tau}_{i+\frac{1}{2}}^x \rangle)^2 \hat{\sigma}_{i}^x \hat{\sigma}_{j}^x  )$,
which is diagonal in the $\hat{\sigma}_{i}^x\hat{\sigma}_{j}^x$ eigenbasis. The
eigenvalues are $\lambda_{1,2}=(1/4)(1 \pm \langle \hat{\tau}_{i-\frac{1}{2}}^x
\hat{\tau}_{i+\frac{1}{2}}^x \rangle)^2$ and $\lambda_{3,4}=(1/4)(1 - (\langle
\hat{\tau}_{i-\frac{1}{2}}^x \hat{\tau}_{i+\frac{1}{2}}^x \rangle)^2)$. Taking derivatives
of the eigenvalues $\lambda_{1,2,3,4}$ and inserting them into Eq.~(\ref{rfsder})
shows that also the two-site fidelity susceptibility diverges as $\chi_F \sim \ln^2 |g/h
-1|$ at $h/g = 1$. Contrary to the case of the Castelnovo-Chamon model, $\chi_F$ for one and two spins now diverges slower than the global fidelity susceptibility, which shows the $\chi_F \sim |g/h - 1|^{-1}$ divergence of the quantum Ising chain~\cite{chen}.

%\section{The Kitaev toric code model in magnetic field}
{\em The Kitaev toric code model in a magnetic field $-$}
Adding a magnetic field $h$ to the Kitaev toric code model~\cite{kitaevtoric} 
gives the Hamiltonian~\cite{trebst}
\begin{equation} \label{toricmhamiltonian}
H=-\lambda_0 \displaystyle \sum_p B_p -\lambda_1 \sum_s A_s - h \sum_i
\hat{\sigma}^{x}_i,
\end{equation}
where the operators $B_p$ and $A_s$ are the same as in Eq.~(\ref{cchamiltonian}). In
the limit $\lambda_1 \gg \lambda_0 , h$, the ground state $|GS\rangle$ will obey
$A_s|GS\rangle = |GS\rangle$, $\forall s$. Then there is a mapping to spin-1/2
operators $\hat{\tau}$ acting on spins at the centers of the plaquettes, according
to $B_p \mapsto \hat{\tau}^x_p$, $\hat{\sigma}_i^x \mapsto \hat{\tau}^z_p
\hat{\tau}^z_q$. Here $i$ is the site shared by the two adjacent plaquettes $\langle
p,q\rangle$. This maps the Hamiltonian (\ref{toricmhamiltonian}) onto~\cite{trebst}
\begin{equation} \label{2dqimhamiltonian}
H=-\lambda_0 \sum_p \hat{\tau}^x_p - h \sum_{\langle p,q \rangle} \hat{\tau}^z_p
\hat{\tau}^z_q,
\end{equation}
which is the 2D transverse field Ising model with magnetic field $\lambda_0 / h =
h'$. Now, the mapping tells us that
$\langle\hat{\sigma}_i^x\rangle = \langle \hat{\tau}^z_p \hat{\tau}^z_q \rangle$,
and the symmetries of the Hamiltonian (\ref{toricmhamiltonian}) imply
$\langle\hat{\sigma}_i^y\rangle = 0$ and $\langle\hat{\sigma}_i^z\rangle = 0$. The
single-site reduced density matrix is therefore given by $\hat{\rho}_i = (1/2)
( 1 +  \langle \hat{\tau}_{p}^z \hat{\tau}_{q}^z \rangle \hat{\sigma}_{i}^x )$,
which has the same form as in the transverse Wen-plaquette model. Since numerical
results have shown a kink in $\langle \hat{\tau}_{p}^z \hat{\tau}_{q}^z \rangle$ at
the phase transition at $h'_c \approx 3$~\cite{trebst}, it follows that the
single-site fidelity will have a drop at this point. Further, the divergence of
$\partial_{h'}\langle \hat{\tau}_{p}^z \hat{\tau}_{q}^z \rangle$ at the critical
point implies a divergence of the single-site fidelity susceptibility at $h'_c$. 
Thus, the scenario that emerges is similar to those for the models above.

\indent {\em Discussion $-$} To summarize, we have analyzed the reduced fidelity at several lattice system TQPTs and found that it serves as an accurate marker of the transitions. In the case of the Castelnovo-Chamon model \cite{castelnovochamon}, the divergence of the reduced fidelity susceptibility at criticality can explicitly be shown to be even stronger than that of the global fidelity \cite{abasto}. Our analytical results rely on exact mappings of the
TQPTs onto ordinary symmetry-breaking phase transitions. Other lattice models exhibiting TQPTs have also been shown to be
dual to spin- \cite{Feng} or vertex \cite{Zhou} models with classical order, suggesting
that our line of approach may be applicable also in these cases, and that
the property that a reduced fidelity can detect a TQPT may in fact be
generic. While counterintuitive, considering that the reduced fidelity is a {\em local probe} of the topologically ordered phase, related results have been reported in previous studies. Specifically, in Refs.~\cite{castelnovochamon} and~\cite{trebst}, the authors found that the local magnetization in the Castelnovo-Chamon model and the Kitaev toric code model in a magnetic field, while being continuous and non-vanishing 
across the transition out of topological order, has a singularity in its first derivative. The fact that local quantities can spot a TQPT is conceptually satisfying, as any physical observable is local in nature. Interesting open questions are here how the concept of reduced fidelity can be applied to
TQPTs in more realistic systems, such as the fractional quantum Hall
liquids, and how reduced fidelity susceptibility singularities depend on different topological and classical orders involved in the transitions.

\indent {\em Acknowledgments $-$} We acknowledge the Kavli Institute for Theoretical Physics at UCSB for hospitality during the completion of this work. This research was supported in part by the National Science Foundation under Grant No. PHY05-51164, and by the Swedish Research Council under Grant No. VR-2005-3942.

\end{document}